\documentclass[fleqn,12pt,twoside]{article}
\usepackage[headings]{espcrc1}

\readRCS
$Id: espcrc1.tex,v 1.2 2004/02/24 11:22:11 spepping Exp $
\usepackage{graphicx}

\usepackage[figuresright]{rotating}

\newcommand{ \la }{\langle}
\newcommand{ \ra }{\rangle}

\newcommand{ \rar }{\rightarrow}

\newcommand{ \pT }{$p_{T}$ }

\newcommand{ \dAu }{$d$ + Au }
\newcommand{ \pp }{$p+p$ }

\newcommand{ \AuAu }{Au + Au }
\newcommand{ \sNN }{$\sqrt{s_{NN}}$ }
\newcommand{ \s }{$\sqrt{s}$ }

\title{Open charm production at RHIC}

\author{Xin Dong\address[USTC]{Department of Modern Physics,
        University of Science and Technology of China - USTC, \\
        96 Jinzhai Road, Hefei, Anhui 230026, China}%
}

\runtitle{Open charm production at RHIC} \runauthor{X. Dong}

\begin{document}

\maketitle

\begin{abstract}
Recent experimental measurements on open charm production in
proton-proton, proton (deuteron)-nucleus and nucleus-nucleus
collisions at RHIC are reviewed. A comparison with theoretical
predictions is made. Some unsettled issues in open charm
production call for precise measurements on directly reconstructed
open charm hadrons.
\end{abstract}

\section{Introduction}

The ongoing four experiments at the Relativistic Heavy Ion
Collider (RHIC) are designed to search for and measure the
quark-gluon plasma (QGP), a new state of matter composed of
deconfined, locally thermalized quarks and gluons. The
equilibrated matter is expected to be described by the equation of
state (EoS) with partonic degrees of freedom. The partonic
pressure gradient and the temperature are two important
characteristics within such kind of EoS.

The physics results from the first three-year runs at RHIC
demonstrate that the partonic pressure gradient has been developed
during the system evolution in heavy ion collisions. This has been
illustrated in the ``white papers" from four
experiments~\cite{whitepaper}. To determine the partonic EoS, the
next task is to test the local and early thermalization hypothesis
{\em experimentally}. Heavy quark ($c,b$) is an ideal probe to
this end. Due to its much heavier mass, it requires more
rescatterings to reach the comparable collectivity as light quark
($u,d,s$). If heavy quark collectivity is observed, there must be
even more rescatterings happening among light quarks than
expected, because the rescattering cross section among light
quarks is larger than that between heavy and light quarks. So
heavy quark collectivity can be used as an indicator for the
thermalization of light flavors, although heavy quarks themselves
do not have to be thermalized~\cite{Dongv2}.

Since charm quark creation requires a large momentum transfer {\it
i.e.} $Q\ ^>_{\sim}$ 3 GeV, it is believed to be less affected by
non-perturbative effects in pQCD calculations. So the measurement
on charm quark production in proton-proton ($p+p$) collisions not
only provides a necessary reference for heavy ion collisions, but
also enables us to test the pQCD calculations on both total and
differential cross sections.

In heavy ion collisions, the theoretical calculation shows that
charm quarks are mostly created through initial gluon-gluon
fusions~\cite{charminit}. And since charm quark mass is much
larger than the estimated system temperature, its production is
little influenced by the thermal component. Unlike light quark,
heavy quark mass is dominated by its current quark mass - the mass
originating from the coupling with the electroweak Higgs
field~\cite{quarkmass}. Therefore heavy quark is an ideal
penetrating probe to the rescatterings and thermalization at the
early stage of heavy ion collisions.

The radiative energy loss of charm quark in vacuum is
characterized by the ``dead-cone" effect~\cite{deadcone}. When an
energetic charm quark traverses through the dense medium, it will
interact with surrounding partons. Theoretical calculations
predict that the suppression of the nuclear modification factor
($R_{AA}$) for charm quarks in central nucleus-nucleus (A + A)
collisions is smaller than that of light
quarks~\cite{elossDM,elossWang,elossNA}. Most of these predictions
were made based on the radiative energy loss mechanism and the
medium properties to our knowledge (gluon density {\it etc.}). The
interaction between charm quark with medium can also be reflected
by the charm quark elliptic flow ($v_2$). The coalescence approach
in a thermalized medium shows that charm hadrons may obtain a
finite $v_2$ even if charm quarks have a zero
$v_2$~\cite{v2eGreco}. The charm quark collectivity has been
studied in an AMPT transport model, and the result shows that a
large charm quark interacting cross section is needed to produce
the magnitude of $v_2$ comparable with that of light
quarks~\cite{v2eZhang}. Measurements of the charm quark
collectivity will tell us the degree to which charm quarks
interact with other partons, and then provide us with pivotal
information on the early thermalization of light flavors.

PHENIX and STAR experiments at RHIC have the capability to detect
charms. PHENIX measures open charms through their semi-leptonic
decays: electrons in central arms in mid-rapidity and muons in
muon arms in forward/backward rapidities. STAR can reconstruct
open charm mesons through theirs hadronic decays in the TPC. It
can also identify electrons with the help of other
sub-detectors~\cite{STARcharmQM04,STARcharmPRL}. The advantage of
PHENIX is that it has low budget materials in its inner detectors,
while STAR has a large acceptance in the TPC around the
mid-rapidity.

\section{Charm production in elementary \pp and $d$ + A collisions}

The first reconstruction of open charm hadrons through their
hadronic decays was reported by the STAR Collaboration in the last
Quark Matter conference~\cite{STARcharmQM04} and was recently
published in Ref.~\cite{STARcharmPRL}. The reconstructed charm
hadrons and decay channels in \dAu collisions are $D^{0}\rar
K^{-}\pi^{+}$, $D^{*+}\rar D^{0}\pi^{+}$ and their charge
conjugate channels. The \pT coverage is $p_{T}<3$ GeV/$c$ for
$D^0$ and $1<p_{T}$/(GeV/$c)<6$ for $D^{*+}$ respectively. Event
mixing technique was used to construct the combinatorial
background in the invariant mass spectrum.

STAR also reported the results of non-photonic electrons mostly
from heavy flavor decays~\cite{STARcharmQM04,STARcharmPRL}.
Results from independent analysis using three different electron
identification methods (r$dE/dx$, $dE/dx$+TOF, $dE/dx$+EMC) are in
good agreement in both \pp and \dAu collisions. The non-photonic
electron spectrum is also consistent with that deduced from STAR's
data for open charm mesons. The combined fit of the total charm
cross section for $D^0$ and non-photonic electrons is
$\sigma_{c\bar{c}}^{NN}=1.4\pm0.2\pm0.4$ mb at \sNN = 200
GeV~\cite{STARcharmPRL}. Within errors, the electron spectra in
\pp and \dAu collisions show approximate $N_{bin}$ scaling,
implying no significant nuclear effect in \dAu collisions.

PHENIX reported its spectra of the non-photonic electrons from
charm hadron decays via three independent methods: the cocktail,
the convertor, and the $\gamma-e$
correlation~\cite{PHENIXeppdAu,PHENIXcharmpp}. These methods all
give a consistent result:
$\sigma_{c\bar{c}}^{NN}=0.92\pm0.15\pm0.54$ mb in \pp collisions
compatible with the STAR data. Measurements of centrality
dependence of non-photonic electrons in \dAu collisions also shows
the $N_{bin}$ scaling~\cite{PHENIXeppdAu}.

\begin{figure}[htbp]
\centering\mbox{
\includegraphics[width=0.7\textwidth]{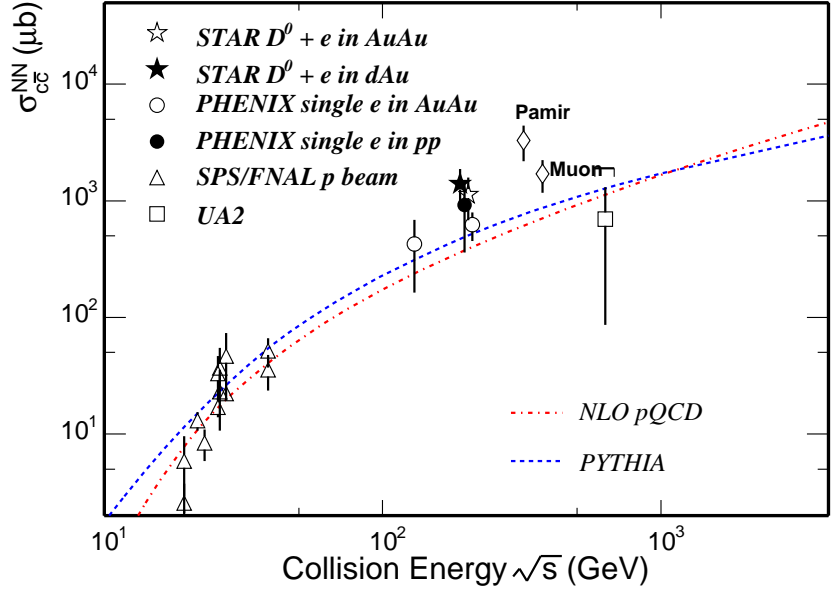}}
\caption{The total $c\bar{c}$ cross section per nucleon-nucleon
collision vs.~the collision energies. The low energy data points
are selected from fixed target
experiments~\cite{lowEcharmreview,DongThesis}. The diamonds depict
two cosmic ray measurements~\cite{cosmic}. The dashed and
dot-dashed lines are taken from~\cite{STARcharmPRL}.}
\label{fig:Xsec}
\end{figure}

Available data for the cross production cross section at various
energies are shown in Fig.~\ref{fig:Xsec}. The results in \AuAu
collisions at RHIC will be discussed in the next section. The
dot-dashed curve depicts a typical next-to-leading (NLO) pQCD
calculation where parameters are optimized to fit the low energy
data~\cite{VogtQCD}. The dashed curve is a PYTHIA (version 6.152)
simulation with the parton distribution function CTEQ5M1. Both the
NLO pQCD calculation and the PYTHIA prediction give a total cross
section of $300-450$ $\mu$b, $2-3$ times lower than the
experimental data. Data points from cosmic ray experiments also
support a large cross section at \s$\sim300$ GeV~\cite{cosmic}.
Recent analysis of open beauty measurements by CDF shows high
order processes ({\it e.g.} initial/final radiation, gluon
splitting, and parton shower production) contribute to a large
part in heavy flavor production at Tevatron~\cite{CDFII}. The
discrepancy at RHIC energy between data and predictions indicates
that these processes may play more important role in theoretical
model than previous thought.

\begin{figure}[htbp]
\centering\mbox{
\includegraphics[width=0.7\textwidth]{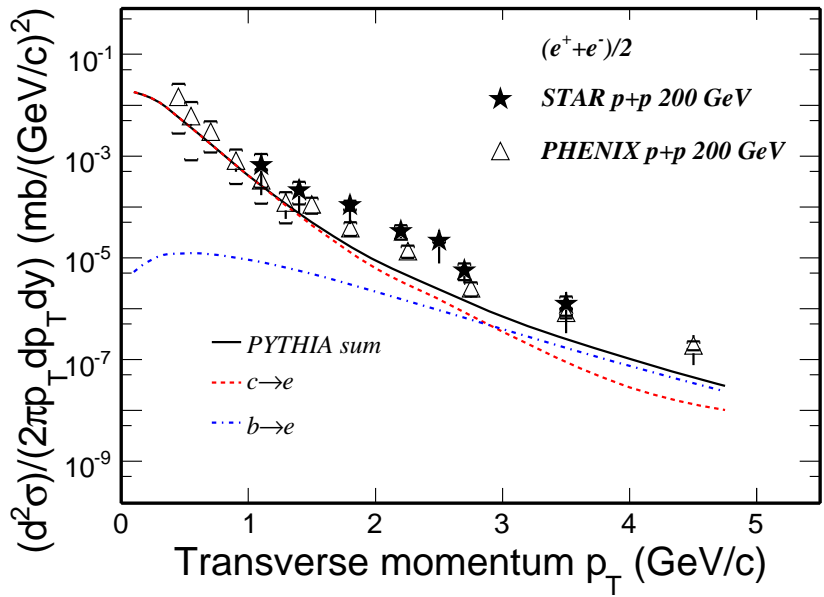}}
\caption{Non-photonic electron spectrum in $p+p$ collisions
compared with PYTHIA model LO calculation. The data points are
taken from ~\cite{STARcharmPRL} and ~\cite{PHENIXcharmpp}. The
PYTHIA model parameter setting is inspired by the publication for
Au + Au 130 GeV data~\cite{PHENIXcharmAuAu130} with
$\sigma_{c\bar{c}}=658$ $\mu$b.} \label{fig:eSpec}
\end{figure}

Apart from an overall normalization factor, the electron spectral
shapes measured by PHENIX and STAR are consistent with each other
within errors. Shown in Fig.~\ref{fig:eSpec}, the measured spectra
are clearly harder than the overall contribution from charm and
bottom decays in the PYTHIA simulation. An important issue in
obtaining the charm production in the electron approach is how to
determine the bottom contamination. A recent NLO pQCD
investigation at RHIC energy tells us that the crossing point
between the electron spectra from charm decays and bottom decays
may vary in a broad \pT range ($\sim3-10$
GeV/c)~\cite{VogtQCDnew}. This will bring very large uncertainties
in electron data from bottom decays.

The reconstructed $D^0$ spectrum covers more than 90\% of the
total charm yields, while the electron spectrum at $p_{T}>0.8$
GeV/$c$ only covers $\sim15$\%. To establish a good reference for
studying charm production in heavy ion collisions, precise
measurements on reconstructed charm hadron with large \pT coverage
are necessary.

\section{Open charm production in A + A collisions}

\subsection{Charm yields}

Charm yields in heavy ion collisions are expected to be scaled by
$N_{bin}$ since most charm quark pairs are created in the initial
hard scatterings. A recent publication from the PHENIX
Collaboration reported the centrality dependence of non-photonic
electron spectra in \AuAu collisions at \sNN = 200
GeV~\cite{PHENIXcharmAuAu200}. The electron spectra in all
centrality bins show approximate $N_{bin}$ scaling with respect to
\pp collisions. The obtained charm total cross section in minimum
bias \AuAu collisions is $0.622\pm0.057\pm0.160$ mb, compatible
with the that in \pp collisions.

At this conference STAR reported reconstructed $D^0$ signals in
minimum bias \AuAu collisions using the same method as that in
\dAu collisions~\cite{STARcharmAuAu}. They also reported
centrality dependence of non-photonic electron measurements in
\AuAu collisions~\cite{STARcharmAuAu,STARcharmAuAuEMC}. By
combining $D^0$ and non-photonic electrons, the extracted total
charm cross section per nucleon-nucleon collision in minimum bias
\AuAu collisions is $\sigma_{c\bar{c}}^{NN}=1.13\pm0.09\pm0.42$
mb, consistent with the STAR \dAu result and the PHENIX \pp
result. The cross section results from \AuAu collisions, measured
by PHENIX and STAR, are illustrated in Fig.~\ref{fig:Xsec}.

\begin{figure}[htbp]
\centering\mbox{
\includegraphics[width=0.7\textwidth]{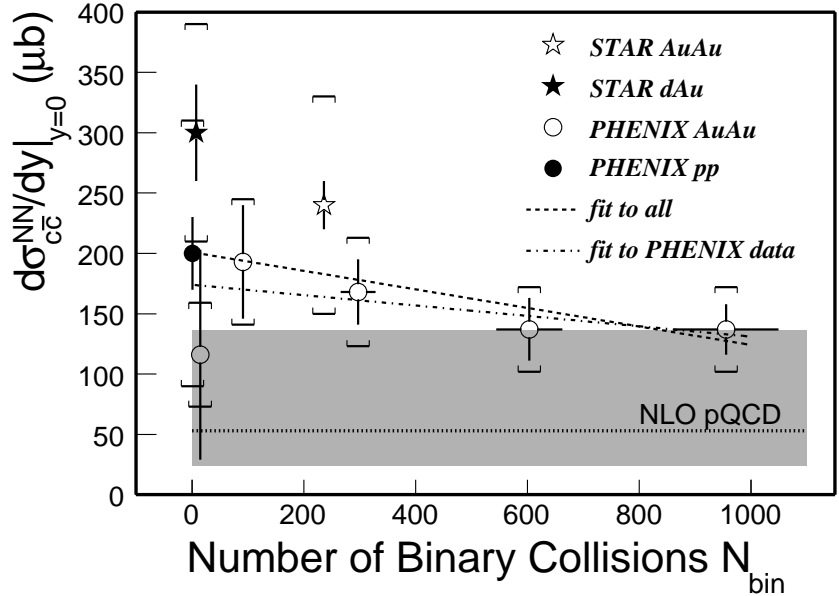}}
\caption{A summary of measurements of the differential $c\bar{c}$
cross section at mid-rapidity per nucleon-nucleon collision vs.
the number of binary collisions at \sNN = 200 GeV. The dotted line
together with the grey band depict a typical NLO pQCD calculation
with uncertainties from Ref.~\cite{VogtQCDnew}. The dashed and
dot-dashed lines are linear fits to all data and PHENIX data
respectively.} \label{fig:XsecNcoll}
\end{figure}

A summary of available data on charm cross sections at 200 GeV is
given in Fig.~\ref{fig:XsecNcoll}. All measurements agree with
each other within large error bars. The dotted line together with
the grey band display a typical NLO pQCD calculation with
uncertainties~\cite{VogtQCDnew}. $N_{bin}$ scaling is assumed in
the extrapolation from \pp to \dAu and to \AuAu collisions. The
data are systematically above this band. One sees from data that
there is a slightly decrease from \pp (\dAu) to peripheral \AuAu,
and then to central \AuAu collisions, as shown by the dashed and
dot-dashed lines which fit to all data and PHENIX ones
respectively. But we cannot claim much due to large errors.

The total cross section measurements are important references for
charmonium production whose enhancement or suppression in central
\AuAu collisions is thought to be a robust signal of the QGP. More
precise charm measurements in various centralities in \AuAu
collisions are needed.

One interesting observation, described in
Ref.~\cite{Xucharmreview}, is that the centrality dependence of
charm yields can be explained marginally by the differential cross
section of inclusive hadrons integrated over $p_T>1.5$ GeV/$c$
$\sim m_{D}$. This means the production of a variety of particles
is not sensitive to the flavor quantity once the momentum transfer
is above the threshold.

\subsection{Charm quark energy loss in medium}

\begin{figure}[htbp]
\centering\mbox{
\includegraphics[width=0.48\textwidth]{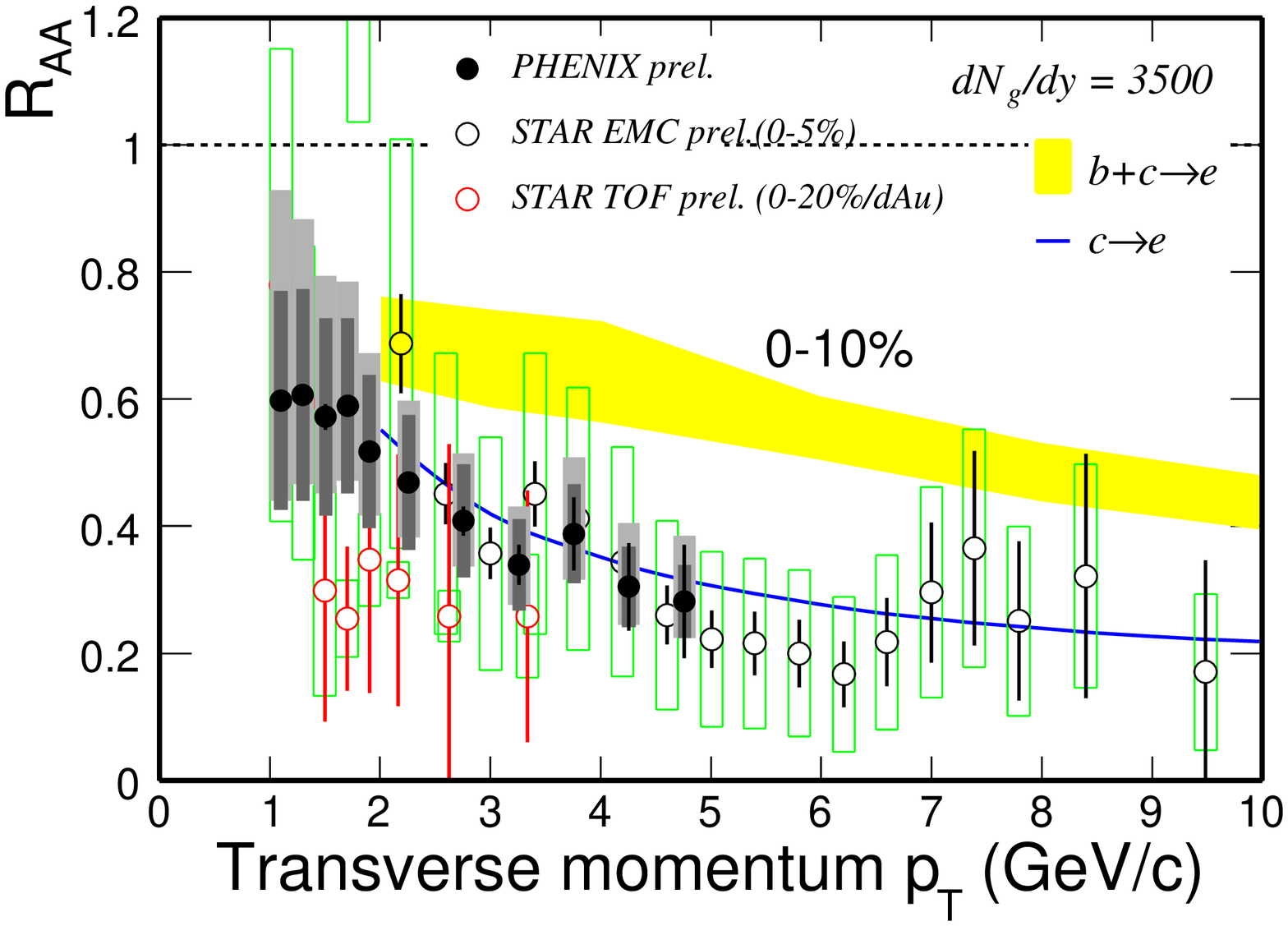}
\includegraphics[width=0.48\textwidth]{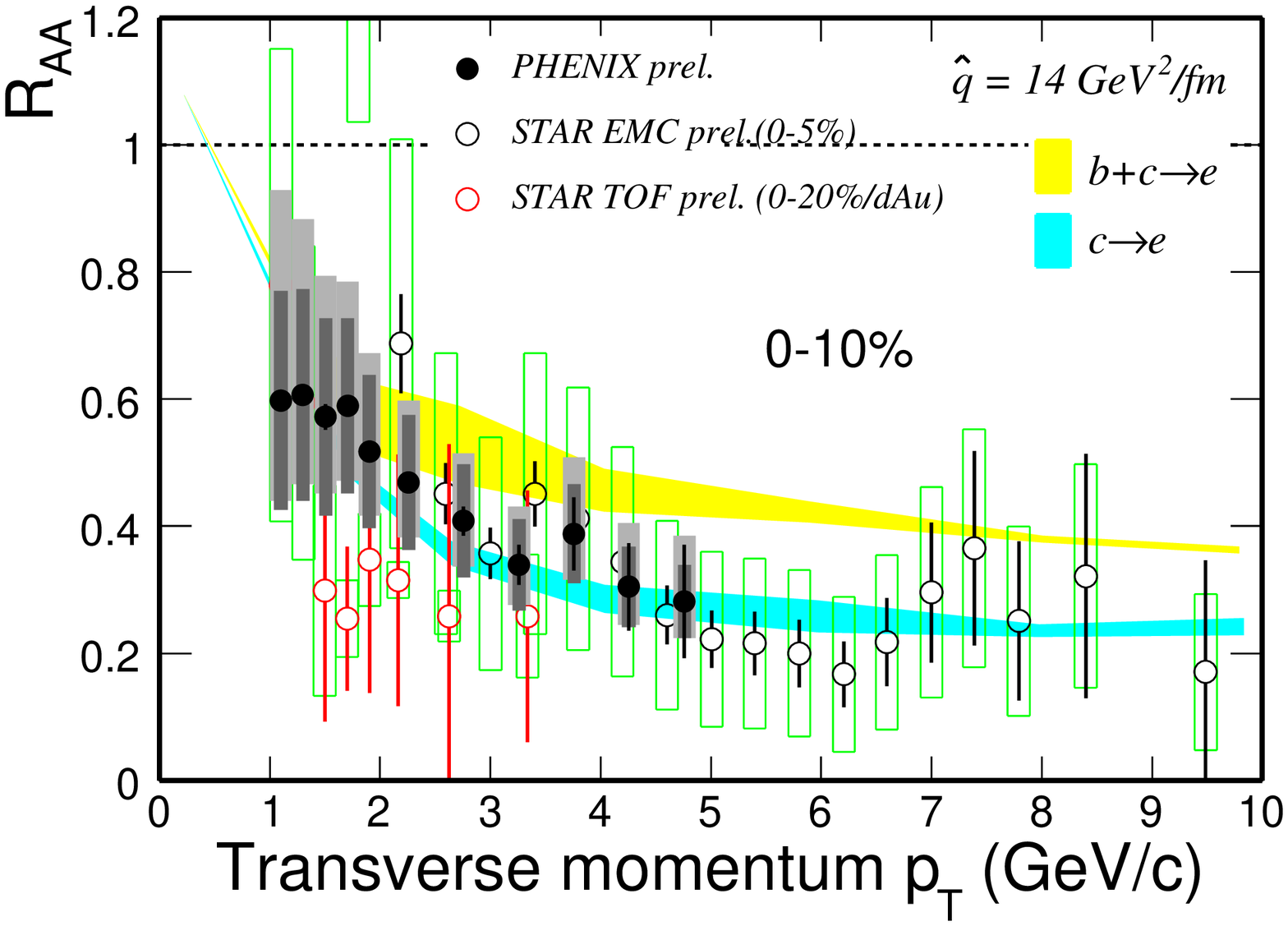}
} \caption{Recent measurements on non-photonic electron $R_{AA}$
in central \AuAu collisions from PHENIX (top 0-10\%) and STAR (top
0-5\% from EMC and top 0-20\% using d+Au as the reference from
TOF) experiments compared with theoretical predictions (top
0-10\%) from ~\cite{elossDMnew} (left plot) and ~\cite{elossNAnew}
(right plot).} \label{fig:RAAe}
\end{figure}

Given in Fig.~\ref{fig:RAAe} are recent results of the nuclear
modification factor $R_{AA}$ of non-photonic electrons in central
\AuAu collisions from PHENIX~\cite{PHENIXeAuAu} and
STAR~\cite{STARcharmAuAu,STARcharmAuAuEMC}. The data give a
consistent, and {\em surprising} fact: the suppression factor for
non-photonic electrons is $\sim0.2-0.3$, which is almost at the
same level as that of charged hadrons in the similar \pT range.
Two recent pQCD estimations in the radiative energy loss scenario
are also shown in that figure~\cite{elossDMnew,elossNAnew}. These
approaches try to fix the transport parameter ($dN_g/dy$ or
$\hat{q}$) boundaries by fitting to the $R_{AA}$ for light hadrons
The boundaries obtained are $1000<dN_g/dy<3500$ and
$4<\hat{q}/$(GeV$^2$/fm)$<14$ respectively. One sees in
Fig.~\ref{fig:RAAe} the upper limit to which energetic partons
lose the largest fraction of their energies in the medium due to
gluon bremsstrahlung from these two approaches. The comparison
with the data illustrates the suppression of electrons from charm
decays may reach as low as that of light hadrons. However, if the
bottom contribution is included according to pQCD calculations,
the overall electron $R_{AA}$ will increase to $\sim0.4-0.5$. This
is a significant discrepancy compared to the data at
$4<p_T$/(GeV/$c)<7$. If the data in Fig.~\ref{fig:RAAe} are
confirmed to be correct, this will bring at least two open issues:
(i) if the current radiative energy loss mechanism persists, there
is no much room for the bottom's contribution in the non-photonic
electron spectrum up to $p_T\sim7$ GeV/$c$. (ii) if the bottom's
contribution is as what is given by the generic pQCD predictions
(the crossing point is $\sim3-5$ GeV/$c$), there must be other
energy loss effects besides gluon radiation. These are challenges
to theorists.

In several recent publications, some authors argued that because
in momentum coverage $\gamma v\sim1$, heavy quark is not
ultrarelativistic, elastic collisional energy loss may play an
important role when charm quarks traverse the
medium~\cite{bothDT,RappReson}. They computed the $R_{AA}$ in the
hydrodynamic transport scenario, which gives the strong
suppression as observed. The approach sheds light on the solution
to the present discrepancy. To decouple the above two issues, one
should precisely measure reconstructed open charm hadrons instead
of electrons. Certainly, a precise reference in \pp collisions on
reconstructed open charm hadrons is needed.

\begin{figure}[htbp]
\centering\mbox{
\includegraphics[width=0.7\textwidth]{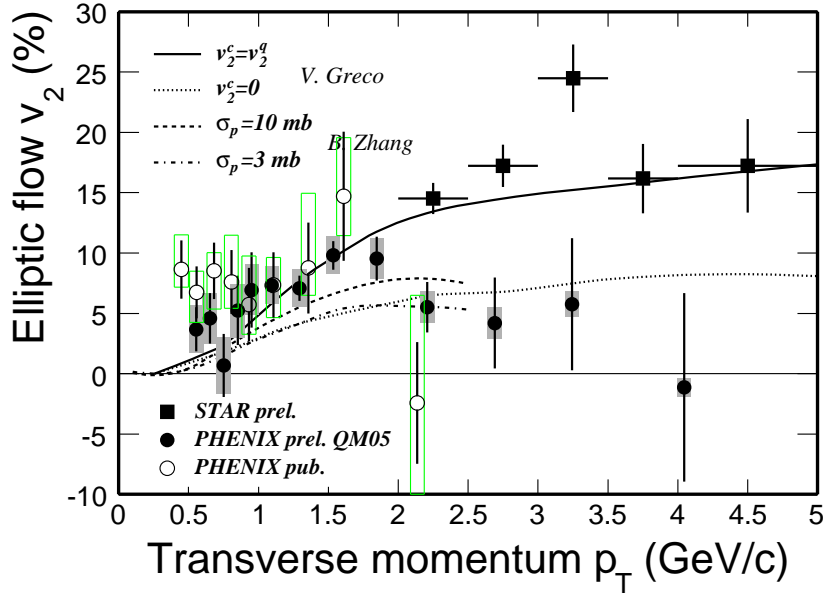}
} \caption{Recent measurements of $v_2$ for non-photonic electron
in minimum bias \AuAu collisions from
PHENIX~\cite{PHENIXeAuAu,PHENIXv2e} and STAR~\cite{STARv2e} and
the comparison with theoretical predictions in
~\cite{v2eGreco,v2eZhang}} \label{fig:ev2}
\end{figure}

\subsection{Charm quark elliptic flow}

PHENIX and STAR recently measured $v_2$ for non-photonic electrons
in minimum bias \AuAu
collisions~\cite{PHENIXeAuAu,PHENIXv2e,STARv2e}, see
Fig.~\ref{fig:ev2}. At $p_{T}>2$ GeV/$c$, even with the claimed
$\sim20-30\%$ systematic error by STAR, the measurements are not
quite consistent between two experiments. In terms of the
magnitude, $v_2$ for non-photonic electrons is comparable to that
of other hadrons~\cite{PIDv2} in $p_T<2$ GeV/$c$. Two model
predictions are also shown in that figure. Since there is an
inconsistency between two experiments and the bottom's
contribution is uncertain at $p_T>2$ GeV/$c$, let us focus on the
range at $p_T<2$ GeV/$c$. In the coalescence model for a
thermalized system~\cite{v2eGreco}, the picture that charm quark
$v_2$ is the same as light quark $v_2$ is supported by the data.
Compared with the transport model results~\cite{v2eZhang}, the
data favors charm quark has a large rescattering cross section,
indicating that the charm quark has a finite $v_2$. Although it is
hard to extract $v_2$ of charm hadrons from electron $v_2$, the
data suggest there can be a non-zero $v_2$ for the charm quark. If
the non-zero charm quark collectivity is confirmed, it means light
flavor thermalization at RHIC, as I have argued in the
introduction section.

\begin{figure}[htbp]
\centering\mbox{
\includegraphics[width=0.7\textwidth]{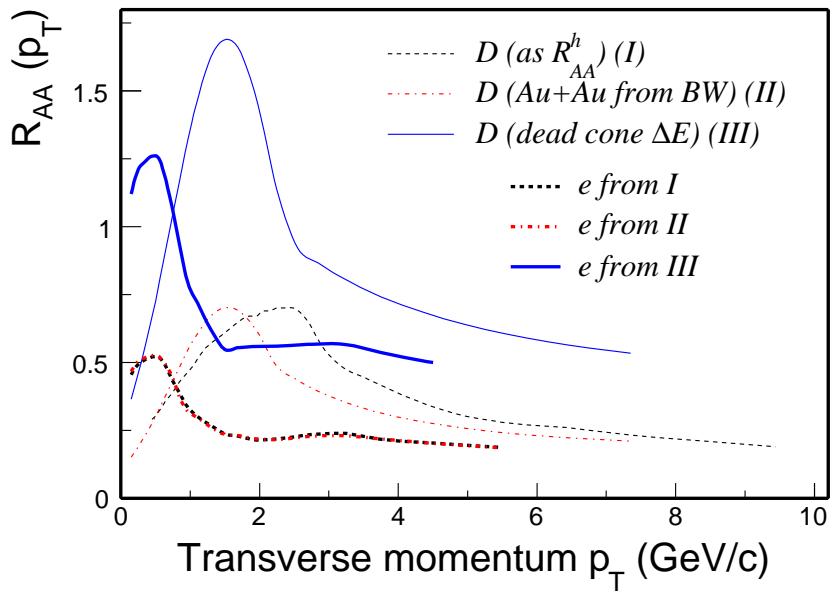}
} \caption{$R_{AA}$ of charm hadrons (thin lines) and decayed
electrons (thick lines) for several assumed $D$ spectra in central
\AuAu collisions. The charm hadron spectrum reference used in \pp
collisions is from STAR's \dAu measurements~\cite{STARcharmQM04}.
The thin dashed line depicts the $R_{AA}$ of charm hadrons same as
that of charged hadrons. The thin dot-dashed line is for charm
hadrons with the Blast-Wave (BW) behavior at $T_{fo}=160$ MeV and
$\la\beta_{T}\ra=0.4c$ in low \pT and with the same behavior as
charged hadrons' $R_{AA}$ in high \pT\cite{whitepaper}. The thin
solid line is for $D$ with the same BW parameters as the
dot-dashed line in low \pT, but the total yield of charm hadrons
is assumed to obey $N_{bin}$ scaling. While in high $p_T$, their
$R_{AA}$ is taken from the dead-cone energy loss calculation in
Ref.~\cite{elossDM}. Thick lines depict the $R_{AA}$ of electrons
from charm decays with the distribution of the same line style.}
\label{fig:RAAce}
\end{figure}

\subsection{Complementary remarks}

From the measurements of $R_{AA}$ and $v_2$ for other particles,
the radiative energy loss can contribute to part of $v_2$, but not
much. In the recent work mentioned above~\cite{bothDT}, elastic
collisions may provide a large fraction of energy loss of heavy
quarks when $\gamma v\sim1$. During hydrodynamical evolution, the
flow of underlying medium will influence the heavy quark spectrum
and heavy quark will pick up some flow. In this case, $R_{AA}$ and
$v_2$ are quite correlated, which means charm quark $R_{AA}$ must
be strong suppressed if a large $v_2$ is observed. So the
combination of measurements on the charm quark spectrum and $v_2$
is essential, especially in low \pT region. The above arguments
are based on the hydro assumptions.

To test the medium response to heavy quarks, the low \pT part is
quite relevant. However, the electron spectrum from charm decays
cannot disentangle different shapes in this \pT region due to
smearing of the decay kinematics~\cite{Nagle}. This effect can
also be reflected on the $R_{AA}$ of non-photonic
electrons~\cite{DongThesis}. The simulation results of various
input charm hadron's $R_{AA}$ (thin curves) and the corresponding
$R_{AA}$ for electrons from charm decays (thick curves) are shown
in Fig.~\ref{fig:RAAce}. One sees that although at high $p_T$, the
electron $R_{AA}$ can reflect the suppression of charm hadrons
(solid curves and dashed curves), the electron $R_{AA}$ at low \pT
cannot tell different thermal shapes (dashed curves and dot-dashed
curves): the thick dashed and thick dot-dashed curves are almost
identical. Therefore we need in the future precise measurements of
the spectrum for reconstructed open charm hadrons at low \pT.

\section{Conclusion and outlook}

The heavy flavor program has started at RHIC extensively. Heavy
flavor collectivity is expected to be an ideal probe to the light
flavor thermalization. Plenty of new and surprising results on
open charm production have been presented on this conference.
However, most of present measurements use electrons decayed from
charm hadrons which brings large uncertainties. The electron
approach can be only a placeholder. We need precise measurements
for reconstructed charm hadron spectra and $v_2$ in a wide \pT
range. So the current sub-detector upgrade proposals in pipe for
PHENIX and STAR detectors are very important to this
goal~\cite{STARTOF,STARHFT,PHENIXSVT}. With the upgraded inner
tracking detector, PHENIX and STAR can reconstruct the secondary
vertices of open charm decays with much lower background. We look
forward to more exciting physical results from RHIC with upgraded
detectors in the future.

\section{Acknowledgement}

I would like to thank the conference organizers for inviting me to
present the talk. I appreciate many constructive discussions with
M. Djordjevic, L. Grandchamp, M. Gyulassy, H. Huang, J. Raufeisen,
H.-G. Ritter, K. Schweda, P. Sorensen, A. Tai, R. Vogt, Q. Wang,
X.-N. Wang, N. Xu, Z. Xu and H. Zhang. This work is partially
supported by the National Natural Science Foundation of China
under the Grant No. 10475071.

\end{document}